# Atom-level design strategy for hydrogen evolution reaction of transition metal dichalcogenides catalysts

Sangjin Lee, Sujin Lee, Chaewon Kim and Young-Kyu Han*

Department of Energy and Materials Engineering, Dongguk University-Seoul, Seoul, 04620, Republic of Korea

*Corresponding Author

E-mail: ykenergy@dongguk.edu

**Abstract**

Two-dimensional transition metal dichalcogenides are among the most promising materials for water-splitting catalysts. While a variety of methods have been applied to promote the hydrogen evolution reaction on the transition metal dichalcogenides, doping of transition metal heteroatoms have attracted much attention since it provides effective ways to optimize the hydrogen adsorption and $H_2$ generation reactions. Herein, we provide in-depth and systematic analyses on the trends of the free energy of hydrogen adsorption ($\Delta G_{H^*}$), the most well-known descriptor for evaluating hydrogen evolution reaction performance, in the doped transition metal dichalcogenides. Using the total 150 doped transition metal dichalcogenides, we carried out the atom-level analysis on the origin of $\Delta G_{H^*}$ changes upon the transition metal heteroatom doping, and suggest two key factors that govern the hydrogen adsorption process on the doped transition metal dichalcogenides: 1) the changes in the charge of chalcogen atoms where hydrogen atoms adsorbed for the early transition metal doped structures, and 2) the structural deformation energies accompanying in introduced dopants for the late transition metal doped structures. Based on our findings, we interpret from a new perspective how vacancies in the TM-doped TMDs can provide optimal $\Delta G_{H^*}$ in HER. We suggest electrostatic control for early TM doped systems and structural control for late TM doped systems as the effective strategies for the thermoneutral $\Delta G_{H^*}$ in TMD.

**Introduction**

As a sustainable and renewable energy source, hydrogen has attracted tremendous attention and a clean generation of hydrogen has been intensively studied for the last decade. While the demands for hydrogen continue to elevate up to about 24% of the total energy needs until 2050 [1], developing practical and clean hydrogen generation methods from water-splitting is urgent because the hydrogen generation still depends largely on the fossil fuels, producing massive amount of carbon dioxides [2–4]. For hydrogen generation from water, the use of catalyst is considered essential, promoting the hydrogen evolution reaction (HER) with curtailed energy consumption. When designing HER catalysts, the hydrogen adsorption free energy, $\Delta G_{H*}$, has been considered the most decisive descriptor to predict the hydrogen evolution reaction performance [5–14].

Two-dimensional transition metal dichalcogenides (TMDs) are among the most noteworthy catalyst materials for HER because of their earth abundancy, high stability and controllable electronic structures [15–26]. From the fact that Pt catalysts, the most superior catalysts for HER, show the near-zero $\Delta G_{H*}$, extensive research efforts have been carried out to design new catalyst materials with thermoneutral $\Delta G_{H*}$, applying the Sabatier principle in a practical form. In particular, replacing transition metal (TM) of TMD with heteroatoms have been considered promising strategies to optimize the $\Delta G_{H*}$ because doped heteroatoms may promote optimal hydrogen adsorption process and $H_2$ generation, facilitating the controls of the catalytic activity [27–36]. For examples, Co, Zn, Pd, and Pt were used to reduce the $\Delta G_{H*}$ of $MoS_2$ by activating their inert basal plane [30–33], and doping of Co, Ni, Cu, and Zn into $MoSe_2$ dragged the high $\Delta G_{H*}$ of non-doped structure into the near-zero values [34–36]. However, although lots of doped TMDs have been demonstrated for enhanced HERs with optimal $\Delta G_{H*}$, the atomic-scale understanding of electrostatic and structural changes during the hydrogen adsorption and appropriate general rules explaining the overall relation between the dopant TM and the $\Delta G_{H^*}$ have not yet been fully understood.

In this study, we provide the universal trends and underlying atom-level origin observed in the $\Delta G_{H^*}$ on the TM doped TMDs. Among TMD materials, $MoS_2$ has attracted tremendous attention for the last decade as a catalyst for HER and has been extensively studied because of its high stability, low cost, and good performance, and doping of $MoS_2$ has been one of the major topics to enhance the HER in $MoS_2$ [28,37–45]. In particular, we pay attention to the stable $2H-MoS_2$ phase since it can work as a robust water-splitting catalyst compared to other metastable phases. From the electronic and structural analysis using first-principles calculations on the 3*d*, 4*d* and 5*d* TM doped $MoS_2$ catalysts, we found that there was a distinctive difference between the early and the late TM dopants with respect to the position of Fermi levels on their *d* bands. While the group 8–12 dopants were expected to show an optimal HER performance with the near-zero $\Delta G_{H^*}$, we revealed that the hydrogen adsorption process in those late TM doped $MoS_2$ is mainly governed by the structural deformation near the dopant. On the other hand, in the early TM doped $MoS_2$, a structural deformation is hardly observed but the change in the charge of the chalcogen atom plays a key role to regulate the hydrogen adsorption. We also clarified that both trends were universally valid for other TMD systems ($MoSe_2$, $MoTe_2$, $CrS_2$, and $WS_2$), where transition metal and chalcogen elements vary. Furthermore, we adapted these findings to verify the role of vacancies in doped TMDs for the optimal $\Delta G_{H^*}$. The chalcogen vacancies in the doped TMDs preliminarily introduce the structural deformation near the dopant even without hydrogen adsorption, then weak Coulomb interaction between adsorbed hydrogen and the late TM dopant leads to the optimal near-zero $\Delta G_{H^*}$. Based on our findings, we provide electrostatic and structural controls as two key strategies for optimal ΔG$_{H*}$, according to the d orbital occupation of metal elements in TMD catalysts, opening a new avenue for improved water-splitting efficiency.

**Method**

Spin polarized ab initio calculations were performed using density functional theory (DFT) provided in Vienna Ab initio Simulation Package (VASP) [46]. We adopted the Perdew-Burke-Ernzerhof exchange-correlation functionals [47] and projector augmented wave method [48]. The 5 × 5 supercell structures were prepared for 2H−MoS$_2$, MoSe$_2$, MoTe$_2$, CrS$_2$, and WS$_2$ monolayers and relaxed until the Hellmann-Feynman forces were less than −0.02 eV/Å. A 600 eV cutoff energy and a 3 × 3 × 1 Monkhorst *k*-point mesh were used for all the computations. The obtained lattice parameters for monolayer MX$_2$ (M = Mo, Cr, and W, and X = S, Se, and Te) were a = 3.18, 3.32, 3.55, 3.04, and 3.19 Å for MoS$_2$, MoSe$_2$, MoTe$_2$, CrS$_2$, and WS$_2$, respectively, which are well agreed with previous reports [49–54]. For those five TMD structures, a TM atom (M) was replaced by a dopant transition metal A (A = 3*d*, 4*d*, and 5*d* transition metals) to describe the TM heteroatom doped TMD structures. The free energy of HER reactions (ΔG$_{H*}$) was computed for all the 150 doped TMD structures using the equation ΔG$_{H*}$ = ΔE$_H$ + 0.24 eV, where ΔE$_H$ is the adsorption energy of a H atom, calculated for 1/2H$_2$ at pH = 0 and *p*(H$_2$) = 1 bar, and the 0.24 eV correction is for the differences in zero-point-energy and entropy [55,56]. A hydrogen atom is attached to a chalcogen atom nearest to the dopant TM. For the cases of the TM doped TMDs with vacancies, a single chalcogen vacancy is placed for the each TMD unit cell at the energetically most stable chalcogen site, which is near the dopant metal atom and a hydrogen atom was adsorbed near the vacancy. The charge for each atom is evaluated using Bader charge analysis [57]. The hydrogen adsorption energy calculated for a hydrogen atom ($\Delta E_{H\_atom}$) is used for the comparison with the structural deformation energy. The structural deformation energy was evaluated using the equation $\Delta E_S = E_0(TMD) - E_S(TMD)$, where $E_0(TMD)$ is the energy of the original TMD structure, and $E_S(TMD)$ is computed without ionic relaxation from the doped TMD with an adsorbed hydrogen atom by replacing the dopant atom with the transition metal atom in the original TMD and removing hydrogen [58]. The doping formation energy was evaluated using the equation $\Delta E_f = E(TM:TMD) - E_0(TMD) - \mu_{TM} + \mu_{TMo}$, where $E(TM:TMD)$ is the energy of the TM doped TMD, and $E_0(TMD)$ is the energy of the original TMD structure, and $\mu_{TM}$ is the energy of a TM atom evaluated for its bulk phase, and $\mu_{TMo}$ is the energy of the original TM atom in its bulk phase. All the atomic structures are illustrated using VESTA code [59].

**Results and discussion**

2H structures of $MoS_2$ were prepared with a single TM atom replaced by a dopant TM heteroatom as shown in Figure 1a. We evaluated $\Delta G_{H*}$, describing the free energy in the hydrogen adsorption process. Figure 1b shows the plots of $\Delta G_{H*}$ for all the doped $MoS_2$ categorized by $3d$, $4d$ and $5d$ TM dopants, and the values are listed in Table 1. It is noticeable that there are two evident trends, which include 1) invariant $\Delta G_{H*}$ for early TM dopants (the group 3, 4, and 5 elements), 2) large variation of $\Delta G_{H*}$ for late TM dopants with the maximum at the group 6 dopants, and the minimum near group 10 dopants. Accordingly, the near-zero values of $\Delta G_{H*}$ are observed near the group 9 dopant. It is also noteworthy that $\Delta G_{H*}$ values increase for the group 11 and 12 dopants, approaching zero. These trends in $\Delta G_{H*}$ reaffirm the experimental reports that the late TM doped TMDs showed superior HER performance [27,28,40,45,60–67].

In order to understand the trends in $\Delta G_{H*}$ with respect to the dopant TMs in MoS$_2$, we analyzed the partial density of states (pDOS) of dopant elements as shown in Figure 2. The *d*-orbitals of dopant TMs have energy gaps between the lower energy states and the higher energy states. The split of *d*-orbitals is attributed to the crystal field under a trigonal prismatic chalcogen coordination in the 2H-TMD structures, where *d*-orbitals of TM split into the higher energy orbitals ($d_{xz}$, $d_{yz}$ > $d_{xy}$, $d_{x^2-y^2}$) and the lower energy orbital ($d_{z^2}$) (Figure 2a) [68–70]. For the early TM dopants (group 3, 4 and 5), the Fermi levels lie in the lower energy bands, which originated from $d_{z^2}$ orbitals (figure 2b-d), while more electrons fill the higher energy bands for the late TMs (Figure 2e-k). The position of the Fermi levels can significantly affect the structural change in hydrogen adsorption process since the attached hydrogen atom will perturb the bonds between the dopant TM and the neighboring sulfur atoms. In the case that the Fermi levels are located in the higher energy states originate under the crystal field from neighboring sulfur atoms, the bond structures near the dopant TM undergoes large structural distortion, directly regulated by the change in the electron filling in the high energy *d*-orbitals.

Figure 3a-j clarify the relation between the Fermi level and the structure deformation when a hydrogen atom is adsorbed. For example, Sc, Ti, V, and Cr doped MoS$_2$ show little break of the symmetry in the sulfur atoms near the dopant even when a hydrogen atom adsorbed (Figure 3a-d). On the other hand, the dopant atoms largely deviated from their original position and the triangular symmetry of neighboring sulfur atoms is broken when a hydrogen atom adsorbed in Mn, Fe, Co, Ni, Cu, and Zn doped MoS$_2$. Since the Fermi levels are located in the higher energy bands of these late TM dopants, the adsorbed hydrogen atom will change the electron filling of the higher energy bands, which originate from $d_{xz}$, $d_{yz}$, $d_{xy}$, and $d_{x^2-y^2}$ orbitals, affecting the bond configuration near the TM dopant. The large deformation of the bond symmetry for the change in higher energy bands is rather inevitable from the fact that the orbitals aligned with the bond from neighboring atoms become unstable, leading to up-shifted energy levels. On the contrary, the change in the lower energy bands have little impact on the bond structure near the dopant atom since down-shifted orbitals have less overlaps with neighboring atoms. The breaking of symmetry near the late TM dopants was also experimentally observed in previous studies [27,71,72]. We found that these relations between the location of Fermi levels and the structural deformation near the dopant atom are generally observed for 4$d$ and 5$d$ TM dopants as shown in Figure S1 and Figure S2, respectively.

The difference in the structural deformation between the early and late TM doped MoS$_2$, which is coming from the relative Fermi level position on the TM's $d$ energy states, implies that the hydrogen evolution reactions proceed in a dissimilar manner. As seen in the $\Delta G_{H*}$ curves in Figure 1b, the free energy stays invariant for the early TM dopants while the large variation is observed for the late TM dopants. Figuring out the distinctive behaviors between the early TM dopants and the late TM dopants on $\Delta G_{H*}$ and understanding what factors dominantly determine $\Delta G_{H*}$, we further analyzed the detailed aspect of hydrogen adsorption on the doped MoS$_2$. Based on the relatively large variation of $\Delta G_{H*}$ values for the late TM doped MoS$_2$, we probed a key factor that governing the hydrogen adsorption process near the late TM dopants. Taking the structural deformation observed for all the late TM dopants (Figure 3 and Figure S2) into account, we analyzed the impact of the distortion in the bonds between the dopant and the neighboring chalcogen atoms when a hydrogen atom adsorbed. The distinctive deformation observed in the late TM doped MoS$_2$ includes a pushed-out TM dopant toward opposite direction to the adsorbed hydrogen atom, invoking one lengthened and two shortened dopant-chalcogen bonds. For example, in the Ni doped MoS$_2$, which has the largest structural deformation among 3$d$ dopants, the lengthened and the shortened Ni−S bonds are 3.19 and 2.23

Å, respectively, originally from the 2.41 Å without the adsorbed hydrogen atom (Figure 3h). Such significant structural deviations in the dopant–neighboring atoms may be attributed to the Coulomb repulsive interaction due to the electric charge of the dopant and the adsorbed hydrogen atom. However, the electric charges of the late TM dopants were lower than those of the early transition metal dopants in general, and the charge of the adsorbed hydrogen atoms is somewhat less positive compared to the early TM doped cases (Figure S3), indicating charge interaction is not predominant in HER on the late TM doped TMDs. The lower positive charges in the late TM dopants than those in the early TM dopants may originate from the decreased difference in the electronegativity between the late TM dopants (1.6−2.4) and sulfur (2.6). The electronegativity of early TM is in the range of 1.2−1.6.

Instead, we focused on the structural deformation energies shown in the late TM doped $MoS_2$. Taking the relations between the Fermi level and the asymmetry in the dopant–neighboring bonds from the distorted crystal fields into account, we analyzed how effectively the late TM doped TMDs optimize their structure when a hydrogen atom adsorbed using the deformation energies. Figure 4 shows comparison between the hydrogen adsorption energy ($\Delta E_{H\_atom}$) and the structural deformation energy ($\Delta E_s$) for $3d$, $4d$, and $5d$ TM doped $MoS_2$. Note that $\Delta E_{H\_atom}$ is evaluated for the hydrogen atom, which is shifted from $\Delta G_{H*}$ by the same amount of energy ($\Delta E_{H\_atom} = \Delta G_{H*} - 2.49$ eV) since the difference is only from the reference energy ($H_2$) and the entropic contribution. We adopted $\Delta E_{H\_atom}$ for this analysis to clearly show the relations between the hydrogen adsorption and the structural deformation in the doped TMD catalyst. The structural deformation energy is evaluated as an energy difference between the doped TMD structure with and without the adsorbed hydrogen atom, where the dopants were replaced by the original TMD elements and the adsorbed hydrogen atom was not considered in order to evaluate contributions solely from the structural change. Note that the structural deformation energy is negligible for the early TM doped MoS$_2$ because no significant structural distortion is observed for them. However, very interestingly, the trends of the structural deformation energies agree well with those of the $\Delta E_{H\_atom}$ for the late TM doped MoS$_2$ (Figure 4), indicating that the structural distortion and relaxation predominantly regulate the hydrogen adsorption on the late TM doped MoS$_2$. The strong correlation between the $\Delta E_{H\_atom}$ and $\Delta E_s$ trends can be understood in terms of the extent to which the doped TMD structures are energetically stabilized by the adsorbed hydrogen atom. As shown in Figure 5, the doping formation energy in the late TM doped MoS$_2$ and the structural deformation energy (Figure 4) have noticeably similar trends with the opposite sign, implying that the breaking of the centro-symmetry caused by hydrogen adsorption effectively alleviate the accumulated energy into TMDs from TM heteroatom doping. It should be mentioned that Hg doped MoS$_2$ shows somewhat unstable doped TMD structures where a Hg dopant largely deviates from the TM metal position in the original TMD, leading to the unusually high doping formation energy (Figure S4). The closely correlated trends between the structural deformation and the hydrogen adsorption suggest that $\Delta G_{H*}$ can be readily tweaked by controlling the distortion of the local structure around the dopants.

Understanding a general role of dopant TM heteroatoms in TMD catalysts, we expanded our computations to other TMD materials and checked if the trends in $\Delta G_{H*}$ observed in TM doped MoS$_2$ were generally valid. We evaluated $\Delta G_{H*}$ for MoSe$_2$ and MoTe$_2$ where S in MoS$_2$ changed to the heavier chalcogen elements, and CrS$_2$ and WS$_2$ where Mo in MoS$_2$ changed to the 3$d$ and 5$d$ transition metal in the same group 6. As shown in Figure 6, the trends in $\Delta G_{H*}$ for MoSe$_2$, MoTe$_2$, CrS$_2$, and WS$_2$ are similar to those for MoS$_2$, implicating our discussion for $\Delta G_{H*}$ in the TM doped MoS$_2$ should be valid for the other TMD catalysts. In fact, we observed that the relation between the position of Fermi levels with respect to $d$ orbitals of TM dopants and the symmetry breaking near the dopant when a hydrogen adsorbed was also very similar to the case of $MoS_2$ (Figure S5–S8). Moreover, the analogous trends in the structural deformation energy with the hydrogen adsorption energy shown in the late TM doped $MoS_2$ were also confirmed in the late TM doped $MoSe_2$, $MoTe_2$, $CrS_2$, and $WS_2$ (Figure S9–S12).

On the other hand, for the case of early TM dopants, relatively invariant $\Delta G_{H*}$ values are observed for the case of early TM dopants within each TMD, whereas different TMDs show dissimilar $\Delta G_{H*}$. For examples, $\Delta G_{H*}$ for Sc, Ti, and V doped $MoS_2$ are 0.29, 0.39, and 0.32 eV, respectively, which is substantially lower than those for $MoSe_2$ (0.73, 0.76, and 0.78 eV for Sc, Ti, and V dopant, respectively). The largest $\Delta G_{H*}$ are observed for those early TM dopants in $MoTe_2$ (0.89, 0.84, and 0.94 eV for Sc, Ti, and V dopant, respectively). These differences in $\Delta G_{H*}$ values seem to originate from the chalcogen atoms (S, Se, and Te) where the charge transfer may occur with adsorbed H atoms. It is worth noting that there is a positive correlation between $\Delta G_{H*}$ and the change in charge of the H-adsorbed chalcogen atom, as seen in Figure 7a. Considering negligible structural deformation observed for H adsorption, it can be inferred that the hydrogen adsorption energy for these early TM doped TMDs may mainly result from the charge transfer between hydrogen atom and the chalcogen atom. When a hydrogen atom adsorbed on the early TM doped TMDs in the HER process, charge transfer occurs between the hydrogen atom and TMDs, acting on lower energy states of the dopant TM atom. While changes in the lower energy states of the dopant TM have little impact on the structural deformation, the hydrogen adsorption energy is dominantly affected by the charge re-distribution determined by the electronegativity of chalcogen atoms. It is also noteworthy that there is no noticeable relation between the charge transfer and $\Delta G_{H*}$ in the late TM doped TMDs as plotted in Figure 7b. For the early TM dopant, therefore, electrostatic control near the chalcogen atom where a hydrogen adsorbed is a key strategy for regulating $\Delta G_{H*}$ in TMD catalysts.

Using the key findings governing HER on doped TMDs, the role of vacancy in the doped $MoS_2$ catalysts for HER can be also explained. While introducing vacancies into the TMD catalysts has been considered an effective strategy for optimizing $\Delta G_{H*}$ and enhancing HER performance [25,73–76], extensive analyses in atomic-scale have not been yet provided enough for the $\Delta G_{H*}$ optimization. As shown in the $\Delta G_{H*}$ evaluation for the doped $MoS_2$ with a sulfur vacancy (Figure 8a), most of the late TM doped $MoS_2$ show the optimized $\Delta G_{H*}$ with near-zero values, indicating that well-balanced hydrogen adsorption and desorption process occur in the $MoS_2$ with vacancies in contrast to the case of pristine $MoS_2$ without vacancies where a relatively large variation in $\Delta G_{H*}$ was observed. Note that the early TM doped $MoS_2$ catalysts with sulfur vacancies do not show the near-zero $\Delta G_{H*}$ values, showing the similarly invariant $\Delta G_{H*}$ as in the non-vacancy cases since there is no participation of the structural deformation. In other words, vacancies in the doped $MoS_2$ successfully alleviate the impact of structural deformation upon hydrogen adsorption which critically governed the $\Delta G_{H*}$ in the non-vacancy cases. We confirmed that the deformation of near-dopant structures, which is the breaking of the triangular centro-symmetry, in the doped $MoS_2$ when a hydrogen adsorbed were similarly seen in the doped $MoS_2$ with vacancies before hydrogen adsorption. For example, as seen in Figure 8b, the dopant in the Ni-doped $MoS_2$ is pushed-out even without an adsorbed hydrogen atom when a sulfur vacancy exists. Note that the aspect of TM dopant shift is in a similar manner with Figure 3h, which includes the pushed-out TM dopant leaving two-shortened TM−S bonds and the widened vacancy space. From the role of structural deformation in HER as discussed in the case of no vacancy, it is inferred that the stabilization of the doped $MoS_2$ structures via TM dopant shift occurs preliminarily to the hydrogen adsorption solely by introducing sulfur vacancies. Meanwhile, $\Delta G_{H*}$ in the case of no structural deformation from hydrogen adsorption become sensitive to the charge interactions as analyzed in the early TM doped $MoS_2$ cases. In particular, the relative low charge of late TM dopant and the increased distance between the late TM dopant and the adsorbed hydrogen atom from the structural deformation may result in the weak Coulomb interactions upon hydrogen adsorption, leading to the optimal $\Delta G_{H*}$ values. The Bader charge of the early and late TM dopants are 1.28−1.76 $e$ and 0.10−1.25 $e$ (Figure S13), respectively, and the distance between TM dopant and adsorbed hydrogen atom is 2.34−3.23 Å in late TM dopants, which is much longer than the those in early TM dopants (2.15−2.42 Å) due to the structural deformation from vacancies (Figure S14). As shown in Figure 8c, $\Delta G_{H*}$ and Coulomb interactions are strongly related in the TM doped $MoS_2$ with vacancies, confirming that weak Coulomb interactions correspond to near-zero $\Delta G_{H*}$ in the late TM doped cases. The relations between $\Delta G_{H*}$ and Coulomb interactions in addition to the detailed structural deformation in the late TM doped $MoS_2$ with vacancies were also applicable to other TMDs as shown in Figure S15-S18.

Although we rationalized the role of vacancies in the doped $MoS_2$ for HER using the two governing factors and explained how those factors effectively optimized the $\Delta G_{H*}$ in the late TM doped cases, adopting vacancies into TMDs is not advantageous for catalytic activities in various aspects. It is known that the vacancies in TMDs may reduce the structural and chemical stability of TMD due to the vacancy clustering [77,78] and the oxidation of chalcogen vacancies [79,80], which may lead to the degradation of HER performance. In addition to the estimation of individual methods for the enhanced performance as HER catalysts in atom-level, and assessment of the strategies in terms of the scope of efficacy, our findings can suggest effective ways to achieve the thermoneutral $\Delta G_{H*}$ specifically for both the early and late TM dopants, providing a practical guideline for probing new TMD based catalysts. For the early TM dopants, electrostatic control of the hydrogen adsorption site is the key strategy since the charge transfer in the chalcogen near the dopant regulates the hydrogen adsorption process. Although the single doping of the early TM showed the non-near-zero $\Delta G_{H*}$, tuning the charge of the chalcogen atoms, for example, by increasing the number of dopant TM atoms near the hydrogen adsorption site, doping into the chalcogen sites, adding non-TM atoms that transferring charges on TMDs, and developing heterostructures for charge transfer with TMDs may help optimize the $\Delta G_{H*}$ of TMD materials. On the other hand, for the late TM dopants, local structural control can promote HER providing the near-zero $\Delta G_{H*}$. Since the breaking of the symmetry near the dopant TM regulate the stability of the hydrogen adsorbed TMD structures, the use of substrates adjusting the structural deformation, adopting the strain into the TMD layers for the tailored dopant-chalcogen distances, using composite structures for the controlled deformation from mismatch, and the doping of multiple types of TM atoms for the tuning of the local structural distortion may provide feasible access to the optimal $\Delta G_{H*}$.

**Conclusions**

The HER performance of the 3*d*, 4*d* and 5*d* TM doped TMD was evaluated using $\Delta G_{H*}$ on $MoS_2$ based catalysts. We revealed that the chemical bond structures between dopant and its neighboring chalcogen atoms when a hydrogen atom adsorbed were regulated by the Fermi level position at the dopant *d* bands. In the case of the early TM doped $MoS_2$, $\Delta G_{H*}$ is related to the change in the charge of the chalcogen atom where a hydrogen atom adsorbed, and no structural deformation is observed. On the other hand, the late TM doped $MoS_2$ undergo a significant structural deformation where the displacement of dopant TM determines how effectively the adsorbed hydrogen stabilize the TMD structures. We verified these trends were generally valid in other TMDs such as $MoSe_2$, $MoTe_2$, $CrS_2$, and $WS_2$. Using the TM dopants' role we found, we could provide rational explanation on the role of vacancies for the optimal $\Delta G_{H^*}$ in the late TM doped $MoS_2$. We suggested that the electrostatic and structural controls near the active sites in TMD catalysts play a key role in achieving the thermoneutral $\Delta G_{H^*}$, and the use of those strategies can be expanded to other TMD materials than the TMD systems considered in this work. Using the relative position between the *d*-band energy states of a dopant TM and Fermi level which is determined by the energy states of TM in TMD, it can be decided which control should be adopted for the dopant. If the strategies for the optimal $\Delta G_{H^*}$ is appropriately selected from among electrostatic controls and the structural controls, the efficiency of material development can be maximized. Considering the intensive attention to the TMD materials as a promising HER catalyst, the two key strategies we suggested can help design a new TMD catalysts for the superior water-splitting capability with enhanced HER performance.

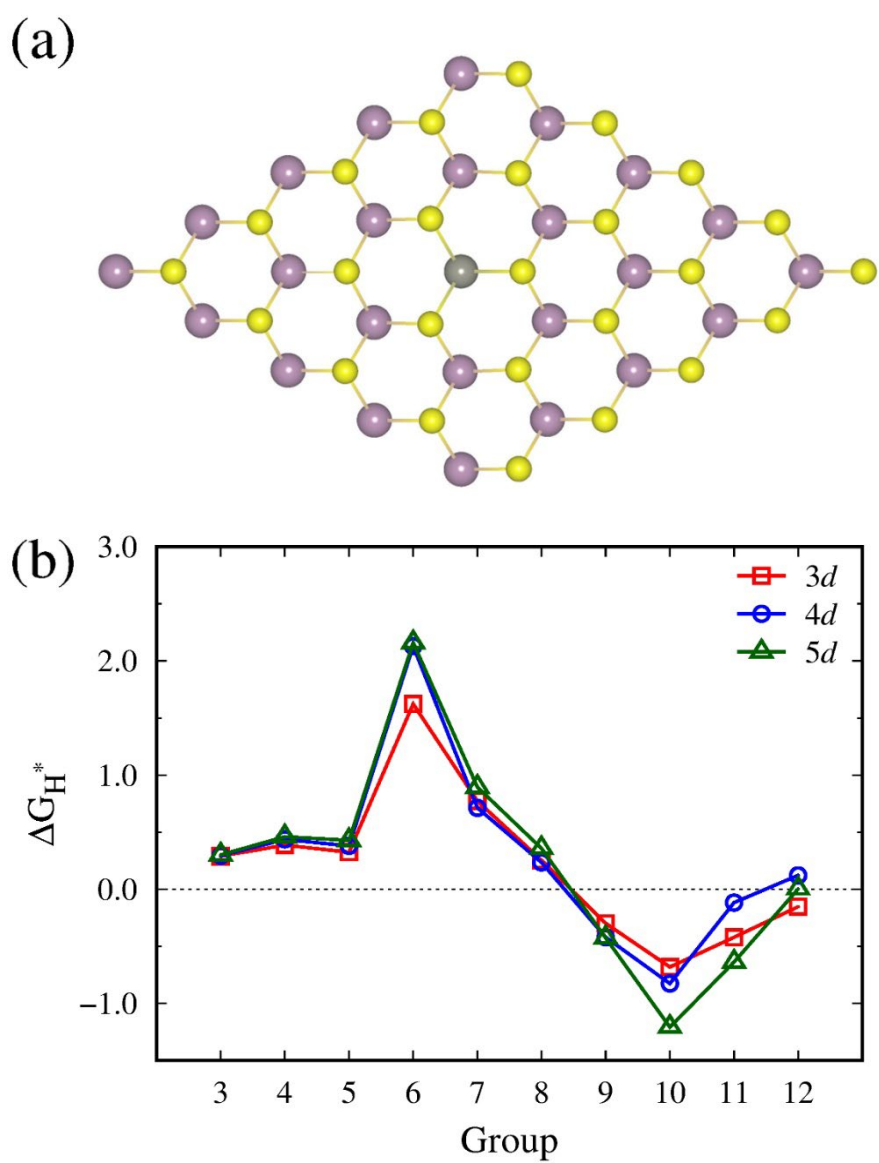

**Figure 1**. (a) An example structure of TM heteroatom doped TMDs, where purple, yellow and grey spheres represent Mo, S and dopant TM, respectively. (b) $\Delta G_{H^*}$ for $3d$, $4d$, and $5d$ TM doped $MoS_2$.

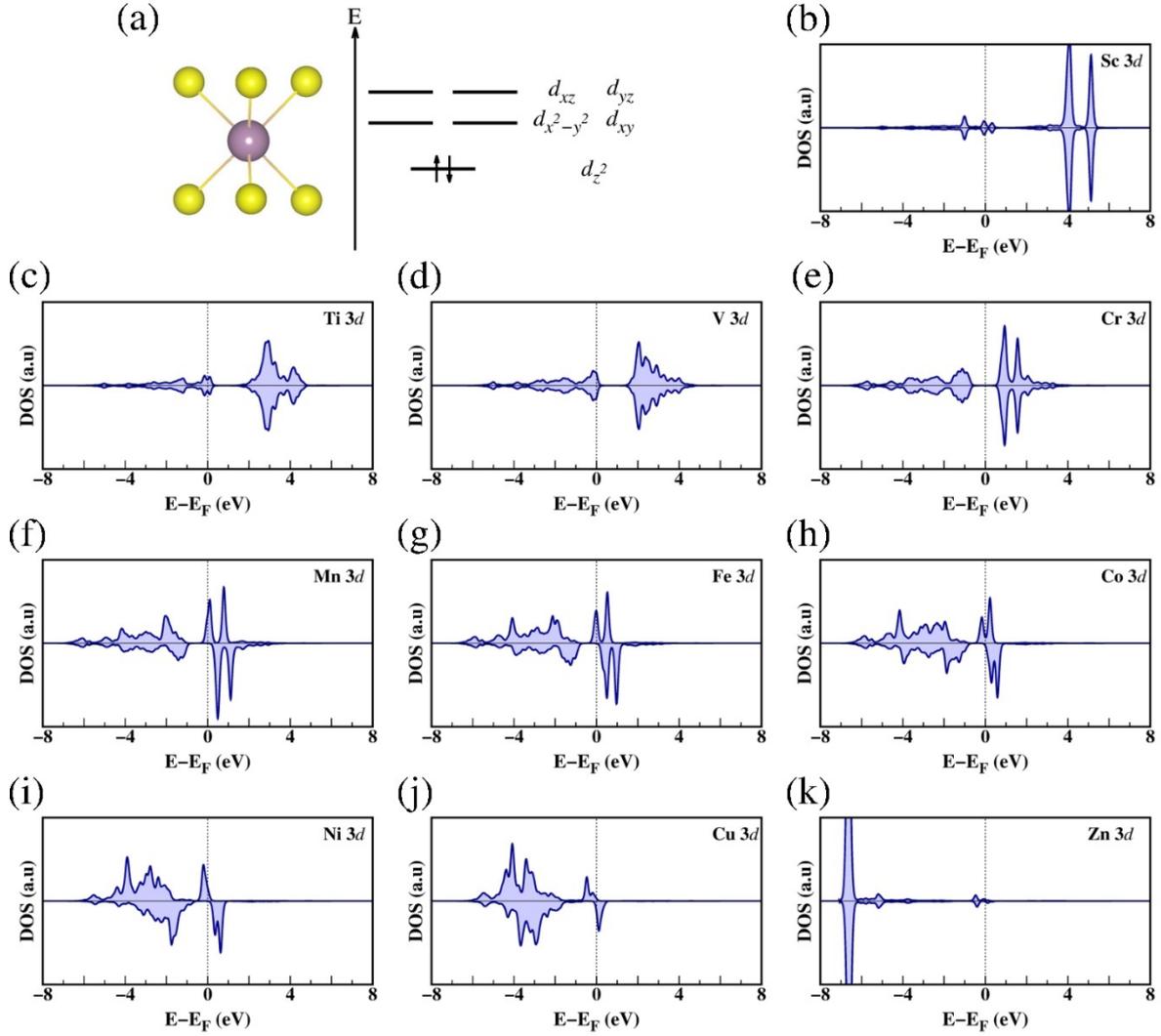

**Figure 2**. (a) A trigonal prismatic structure of 2H-TMDs (left) and *d* band alignment of TM under the crystal field from chalcogen atoms (right). Partial density of states of 3d band of a dopant TM in the TM, (b) Sc, (c) Ti, (d) V, (e) Cr, (f) Mn, (g) Fe, (h) Co, (i) Ni, (j) Cu, and (k) Zn, doped $MoS_2$.

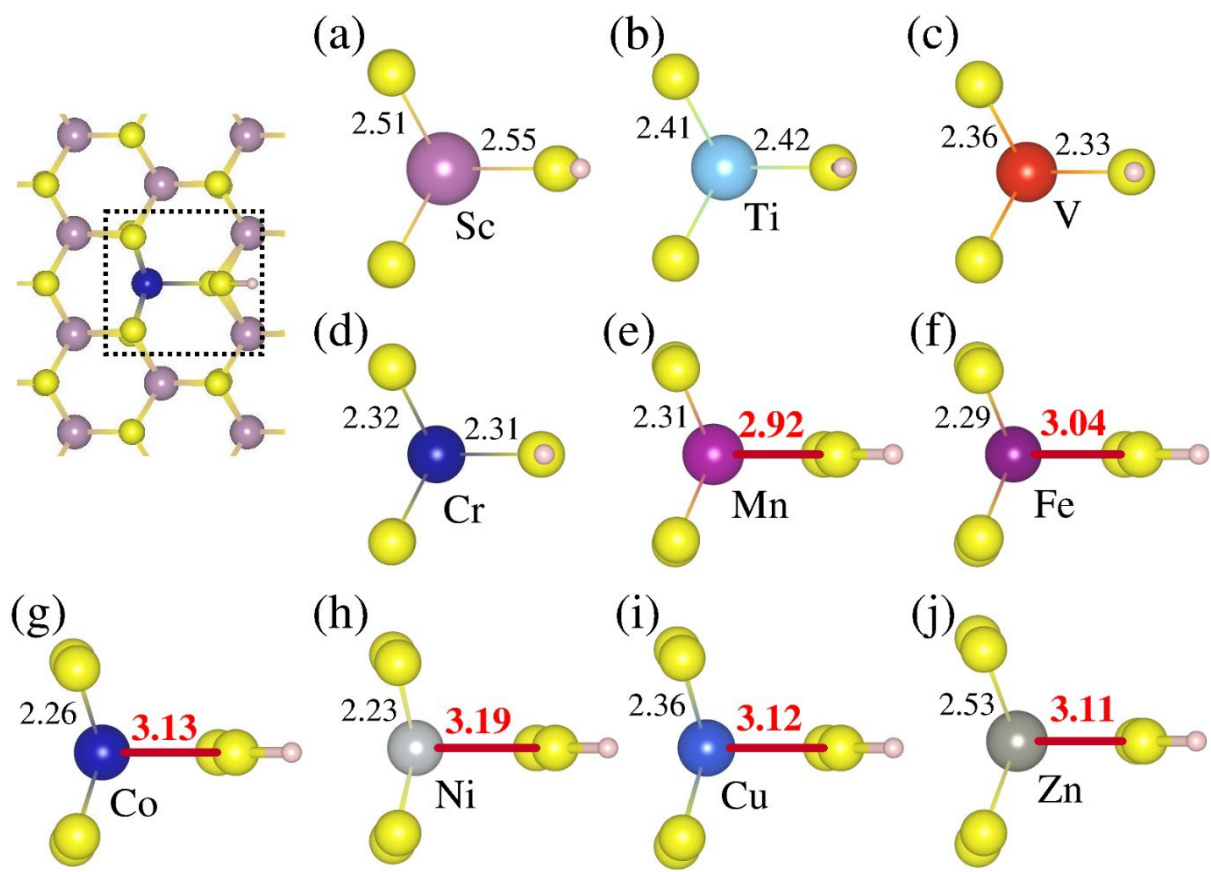

**Figure 3**. Structures of the TM doped MoS$_2$ when a hydrogen atom adsorbed. The purple, yellow, and pink spheres represent Mo, S and H atom, respectively, and the bond lengths (Å) are marked for dopant-chalcogen bonds. The triangular centro-symmetry near the dopant is maintained in early TM doped MoS$_2$ (a-d) and broken in late TM doped MoS$_2$ (e-j).

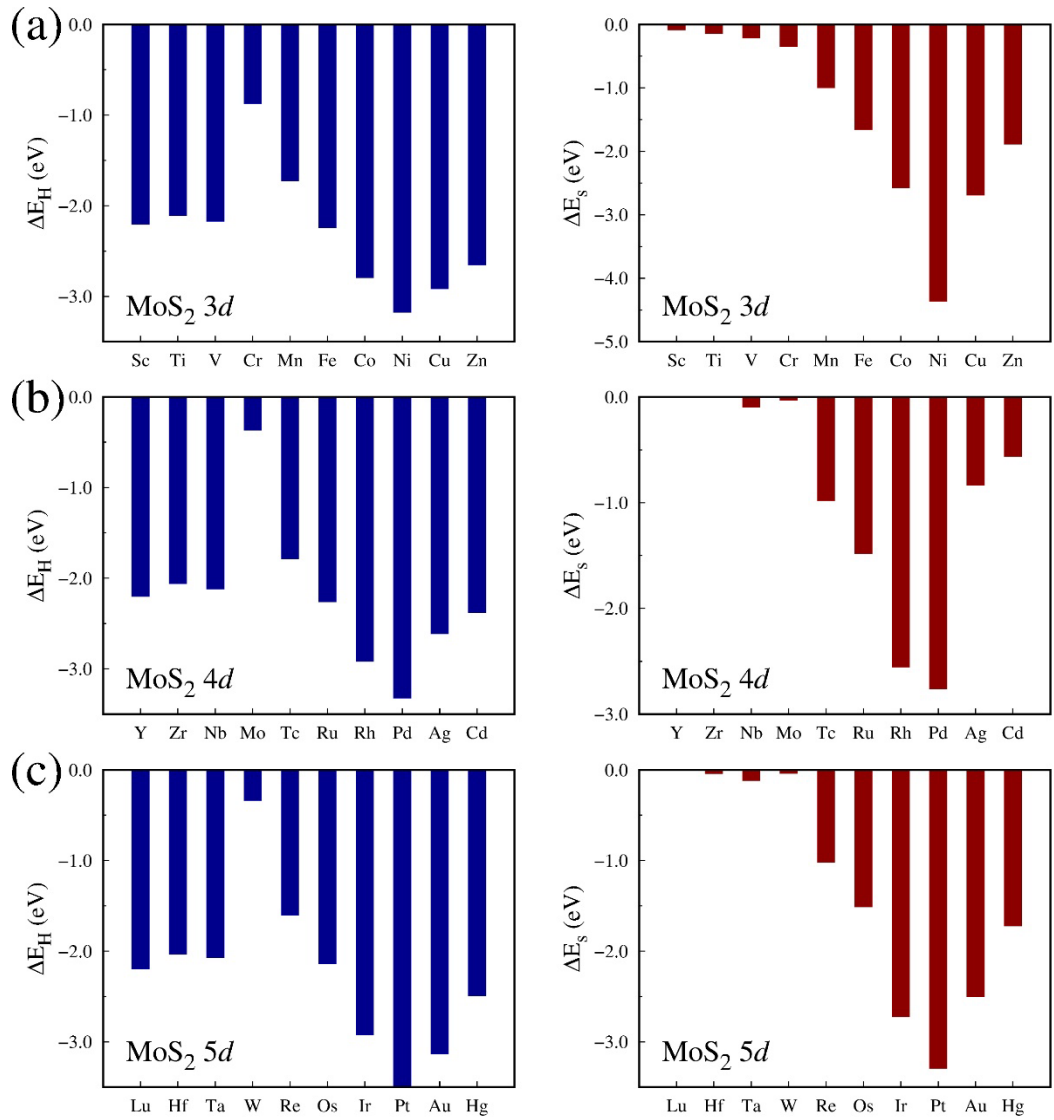

**Figure 4**. Comparisons between the hydrogen adsorption energy ($\Delta E_{H\_atom}$) and the structural deformation energy ($\Delta E_s$) for (a) 3$d$, (b) 4$d$, and (c) 5$d$ TM doped MoS$_2$.

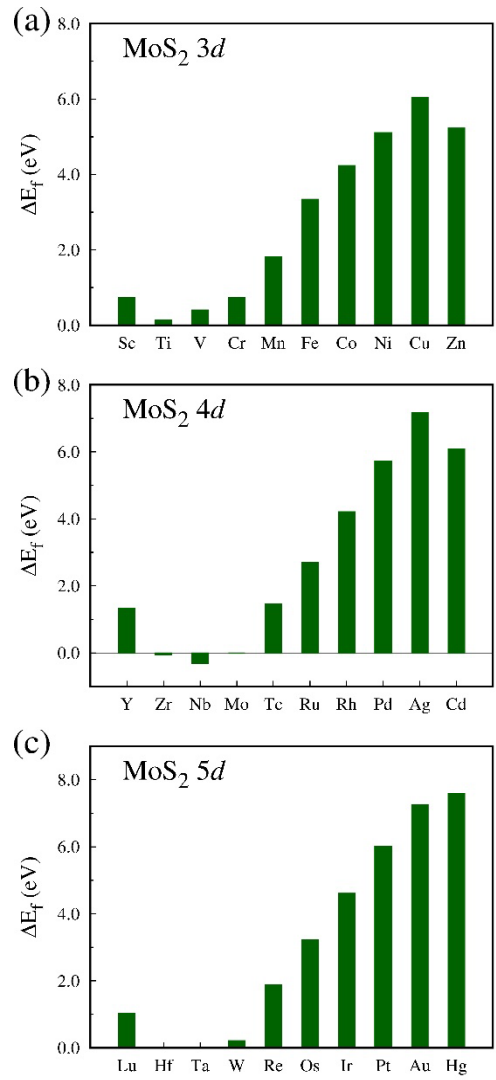

**Figure 5**. The doping formation energy (ΔE$_f$) for (a) 3*d*, (b) 4*d*, and (c) 5*d* TM doped MoS$_2$.

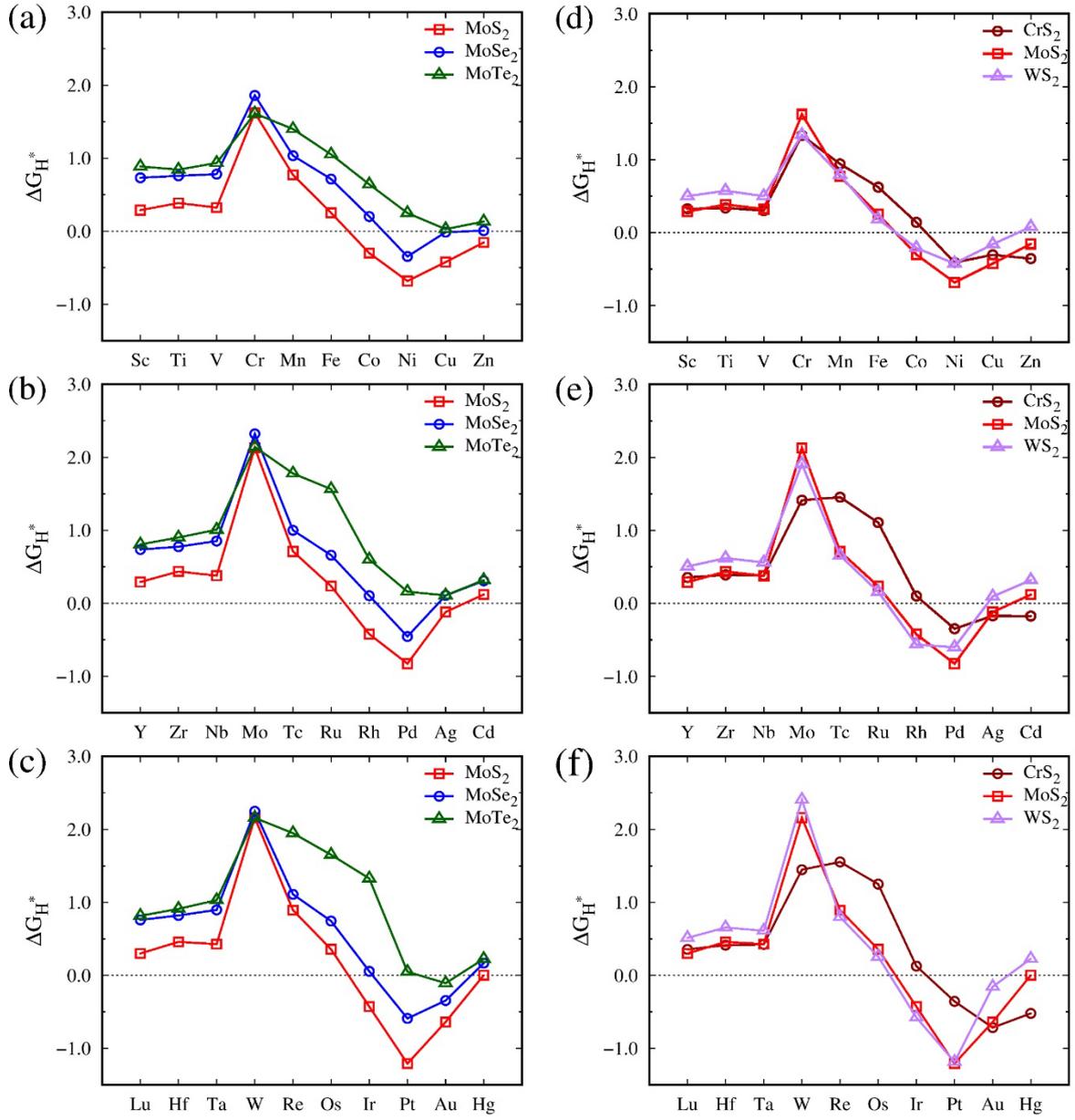

**Figure 6**. ΔG$_{H*}$ for (a) 3*d*, (b) 4*d*, and (c) 5*d* TM doped MoS$_2$, MoSe$_2$, and MoTe$_2$, and ΔG$_{H*}$ for (e) 3*d*, (e) 4*d*, and (f) 5*d* TM doped CrS$_2$, MoS$_2$, and MoTe$_2$.

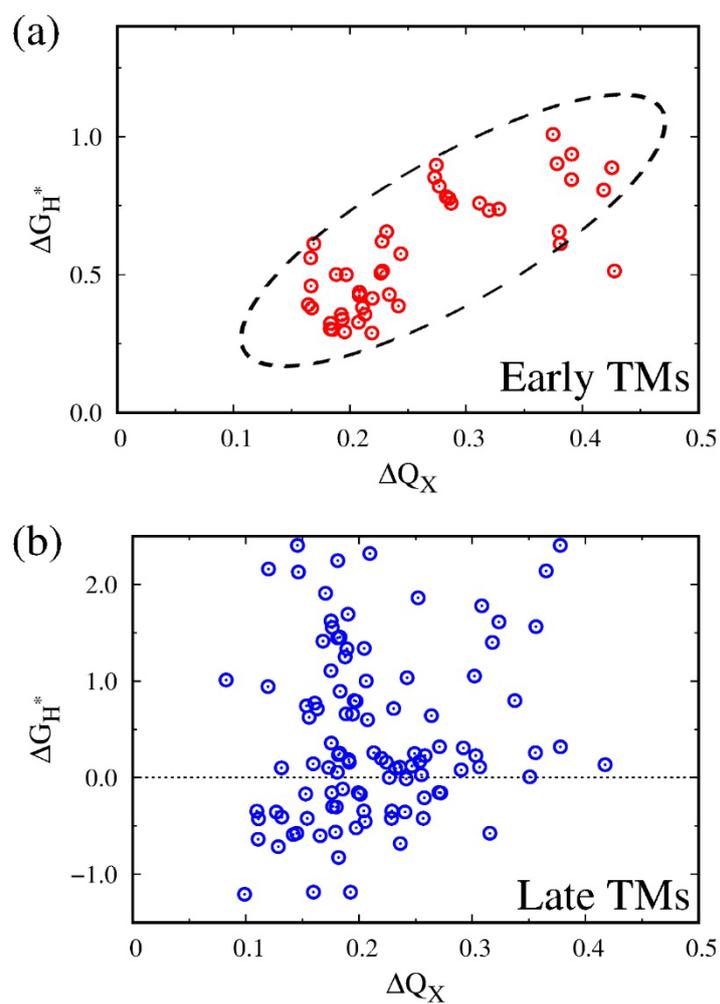

**Figure 7.** Relations between $\Delta G_{H^*}$ and the changes in charge of chalcogen atom when a hydrogen adsorbed ($\Delta Q_X$) for (a) early TM and (b) late TM doped TMDs. $\Delta Q_X$ is the difference in the Bader charges of the chalcogen atom before and after the hydrogen adsorption.

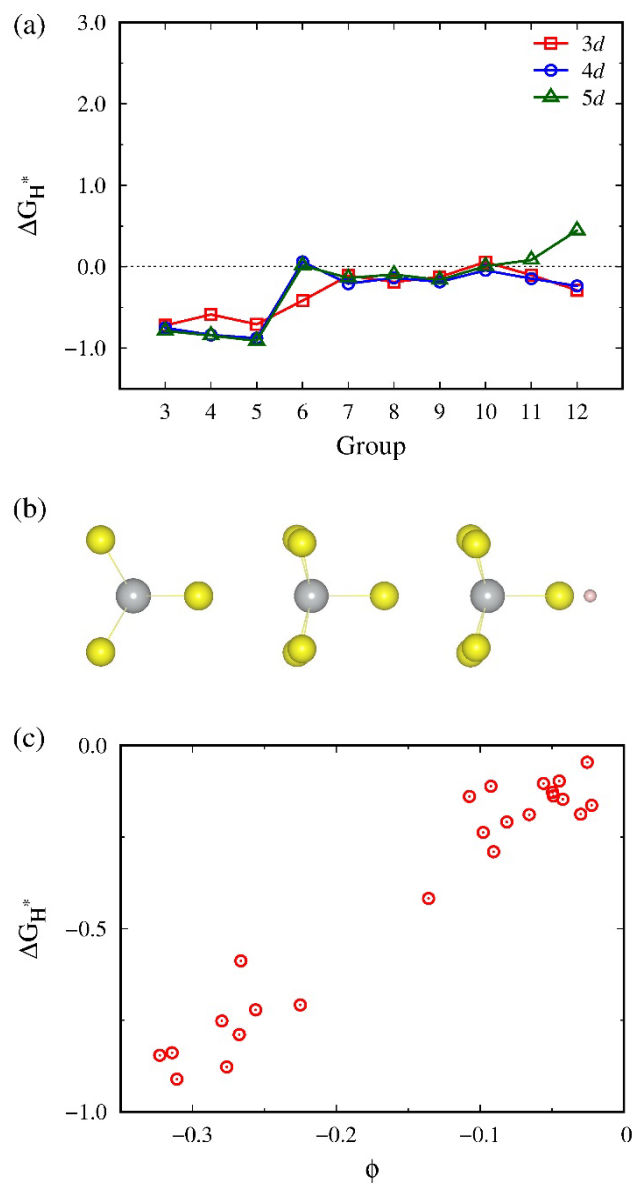

**Figure 8**. (a) $\Delta G_{H*}$ for *3d*, *4d*, and *5d* TM doped $MoS_2$. (b) Structures of Ni doped $MoS_2$ without vacancies (left), with vacancies (center) before hydrogen adsorption, and with vacancies (right) after hydrogen adsorption. The purple, yellow, grey, and pink spheres represent Mo, S, Ni and H atoms, respectively. (c) Relations between $\Delta G_{H*}$ and the Coulomb interaction ($\phi$) between the TM dopant and the adsorbed hydrogen atom. The Coulomb interaction is evaluated using the Bader charges of the dopant and hydrogen atom, and their distance.

**Table 1**. $\Delta G_{H*}$ (eV) of $3d$, $4d$ and $5d$ TM doped $MoS_2$ monolayer structures.

| $3d$ | $\Delta G_{H*}$ | $4d$ | $\Delta G_{H*}$ | $5d$ | $\Delta G_{H*}$ |
|---|---|---|---|---|---|
| Sc | 0.29 | Y | 0.29 | Lu | 0.30 |
| Ti | 0.39 | Zr | 0.44 | Hf | 0.46 |
| V | 0.32 | Nb | 0.38 | Ta | 0.43 |
| Cr | 1.62 | Mo | 2.13 | W | 2.16 |
| Mn | 0.77 | Tc | 0.71 | Re | 0.89 |
| Fe | 0.25 | Ru | 0.24 | Os | 0.36 |
| Co | −0.30 | Rh | −0.42 | Ir | −0.43 |
| Ni | −0.68 | Pd | −0.83 | Pt | −1.21 |
| Cu | −0.42 | Ag | −0.12 | Au | −0.64 |
| Zn | −0.15 | Cd | 0.12 | Hg | 0.00 |

Thong, Selective Engineering of Chalcogen Defects in MoS$_2$ by Low-Energy Helium Plasma, ACS Appl. Mater. Interfaces. 11 (2019) 24404–24411.